\documentclass[iop]{emulateapj}
\slugcomment{Accepted by the Astrophysical Journal}
\shorttitle{X-ray Constraints on VY CMa}
\shortauthors{Montez et al.}
\begin{document}

\title{Constraints on the Surface Magnetic Fields and Age of a Cool Hypergiant: XMM-Newton X-ray Observations of VY CMa}

\author{Rodolfo Montez Jr.}
\affil{Department of Physics and Astronomy, Vanderbilt University, Nashville, TN 37235, USA}

\author{Joel H. Kastner} 
\affil{Center for Imaging Science, School of Physics and Astronomy, and Laboratory for Multiwavelength Astrophysics, Rochester Institute of Technology, 54 Lomb Memorial Drive, Rochester, NY 14623, USA} 

\author{Roberta M. Humphreys}
\affil{Minnesota Institute for Astrophysics, 116 Church Street SE, University of Minnesota, Minneapolis, MN 55455, USA} 

\author{Rebecca Leigh Turok} 
\affil{Vanderbilt University, Nashville, TN 37235, USA}

\author{Kris Davidson}
\affil{Minnesota Institute for Astrophysics, 116 Church Street SE, University of Minnesota, Minneapolis, MN 55455, USA}

\begin{abstract}
The complex circumstellar ejecta of highly evolved, cool hypergiants are indicative of multiple, asymmetric mass loss events. To explore whether such episodic, non-isotropic mass loss may be driven by surface magnetic activity, we have observed the archetypical cool hypergiant VY CMa with the XMM-Newton X-ray satellite observatory.  The hypergiant itself is not detected in these observations. From the upper limit on the X-ray flux from VY CMa at the time of our observations ($F_{X,UL}~\approx~8\times10^{-14} {\rm~erg~cm}^{-2} {\rm~s}^{-1}$, corresponding to $\log~L_X/L_{bol}\leq-8$), we estimate an average surface magnetic field strength $fB \leq 2\times10^{-3}$~G (where $f$ is the filling factor of magnetically active surface regions). These X-ray results for VY CMa represent the most stringent constraints to date on the magnetic field strength near the surface of a hypergiant. VY CMa's mass loss is episodic and may have been in a state of low surface magnetic activity during the XMM observations. The XMM observations also yield detections of more than 100 X-ray sources within $\sim$15$^{\prime}$ of VY CMa, roughly 50 of which have near-infrared counterparts. Analysis of X-ray hardness ratios and IR colors indicates that some of these field sources may be young, late-type stars associated with VY CMa, its adjacent molecular cloud complex, and the young cluster NGC 2362. Further study of the VY CMa field is warranted, given the potential to ascertain the evolutionary timescale of this enigmatic, massive star.
\end{abstract}

\section{Introduction}

High-resolution imaging of the cool, rapidly mass-losing hypergiants\footnote{We use the term hypergiant to refer to evolved stars with spectral types from late A to M that lie just below the upper luminosity boundary in the HR Diagram.} VY CMa and IRC+10420 have yielded evidence for multiple, asymmetric mass loss events \citep{1997AJ....114.2778H,2001AJ....121.1111S,2002AJ....124.1026H,2005AJ....129..492H, 2007AJ....133.2716H,2010AJ....140..339T}.
These stars are the most luminous known for their apparent photospheric temperatures, and their mass loss rates rank among the highest for cool stars (i.e., a few $\times 10^{-4}$ M$_{\odot}$ yr$^{-1}$). 
It has been suggested \citep{2001AJ....121.1111S,2005AJ....129..492H, 2007AJ....133.2716H,2010AJ....140..339T} that the complex loops, arcs and knots in their ejecta are possibly due to strong surface magnetic activity. 
Recent ALMA observations of VY CMa, which reveal distinct, asymmetrically distributed clumps of ejecta on scales of tens of stellar radii, further reinforce such a notion  \citep{2014arXiv1410.1622O}. 

Indeed, the presence of starspots, as manifest by large surface brightness anisotropies, has been inferred for several cool supergiants. 
The best-studied example is  $\alpha$ Ori \citep{1996ApJ...463L..29G}, but stellar hotspots have also been observed on $\alpha$ Sco and $\alpha$ Her with properties consistent with a convective, hence magnetic, origin \citep{1997MNRAS.285..529T}. 
Interferometric  imaging of cool hypergiant stars reveals large-scale inhomogeneities attributed to magnetic fields and/or rotation \citep{2004ApJ...605..436M}.  
Magnetic fields have now been measured in the ejecta of several cool hypergiant stars \citep{2002A&A...394..589V,2005MmSAI..76..462V}, lending support to the notion that their episodic mass loss may be driven by surface magnetic activity.

For rapidly mass-losing evolved stars, the potential connection between such magnetic activity and coronal X-ray emission remains to be established.  
Luminous X-ray emission from coronae associated with luminous red giants and supergiants is rarely observed \citep[][]{2003ApJ...598..610A,2005ApJ...618..493A}. 
However, there exists a dearth of meaningful measurements of the most rapidly mass-losing evolved stars; to assess the X-ray activity levels of such stars requires deep observations with adequate hard ($\gtrsim$1 keV) X-ray sensitivity to penetrate their thick circumstellar envelopes.  

The mass loss asymmetries that develop during the very late stages in the evolution of massive stars are analogous to those that characterize stars on the asymptotic giant branch (AGB), post-AGB stars, and the very youngest planetary nebulae (PNe).  
Only a few such cool, rapidly mass-losing AGB and post-AGB stars have been observed by the two contemporary X-ray observatories that are well-suited to such work (i.e., Chandra and XMM-Newton). 
In XMM observations, the single, mass-losing AGB stars TX Cam and T Cas were not detected \citep{2004ApJ...608..978K} whereas Mira AB --- a wide-separation AGB star and white dwarf (WD) binary that drives a large-scale bipolar outflow \citep{2009A&A...500..827M} --- is quite luminous in X-rays \citep{2004ApJ...616.1188K}. 
Chandra imaging further revealed that both components of the Mira AB binary system are X-ray-luminous, and that Mira A (the AGB star) evidently undergoes energetic flaring  \citep{2005ApJ...623L.137K}.
The two additional known examples of X-ray-emitting AGB stars are also likely binary systems \citep{2012A&A...543A.147R}.
Meanwhile, Chandra observations are establishing that a significant fraction of the central stars of planetary nebulae are X-ray sources; as for AGB stars, these planetary nebula central star X-ray sources are frequently (though not exclusively) associated with binarity \citep{2010ApJ...721.1820M,2012AJ....144...58K,2014arXiv1407.4141F,ChanPlaNSIII}. 

Moving higher on the HR diagram, only a few X-ray observations of supergiants and hypergiants have been obtained with modern X-ray observatories: the iconic red supergiant $\alpha$ Ori  is apparently ``dark'' in X-rays \citep{2006astro.ph..6387P}, and the yellow hypergiant IRC+10420 was not detected by XMM \citep{2014NewA...29...75D}.

Here, we report on a search for X-ray evidence of magnetic activity at VY CMa (HD 58061), via XMM-Newton observations.
VY CMa is a special case even among the cool hypergiants. 
A powerful infrared and maser source, it is one of the most luminous and largest evolved cool stars known, and features one of the highest measured mass loss rates among evolved stars \citep[2-3$\times10^{-4} M_{\odot} {\rm ~yr}^{-1}$;][]{1994AJ....107.1469D}. 
The star is surrounded by a complex, asymmetric dust reflection nebula $10^{\prime\prime}$ in extent  \citep{1972ApJ...172..375H,1998AJ....115.1592K,2001AJ....121.1111S}; this dusty circumstellar material obscures
the central star across the visible wavelength range \citep{1998AJ....115.1592K}. 
Visual imaging, spectroscopy, and polarimetry \citep{2001AJ....121.1111S,2005AJ....129..492H, 2007AJ....133.2716H,2007AJ....133.2730J} reveal that the nebula's expanding arcs, prominence-like loops and knots were ejected at different times and in different directions, apparently by localized processes from different regions on the star.  

As argued in \citet{2002A&A...394..589V}, measurements of  polarization from masers in the ejecta of VY CMa appear to trace a magnetic field that increase with strength moving closer to the surface of VY CMa.
At large distances from the star ($\sim 2800$ AU), OH maser polarization suggests a field strength of $\sim 2$ mG  \citep{1987MNRAS.225..491C};  at $\sim 200$ AU, H$_2$O maser polarization suggests 200 mG \citep{2002A&A...394..589V}; and, nearest the star, SiO maser polarization suggests a field strength of 65 G \citep{1987Natur.329..613B}. 
This trend in the magnetic field strength from maser polarization is suggestive of a global field strength that drops like r$^{-2}$.
Extrapolating this radial dependence implies that magnetic field strengths on the order of $10^{2}$ to $10^{3}~{\rm G}$ may exist at the surface of VY CMa.

\begin{figure*}
\includegraphics[scale=0.75]{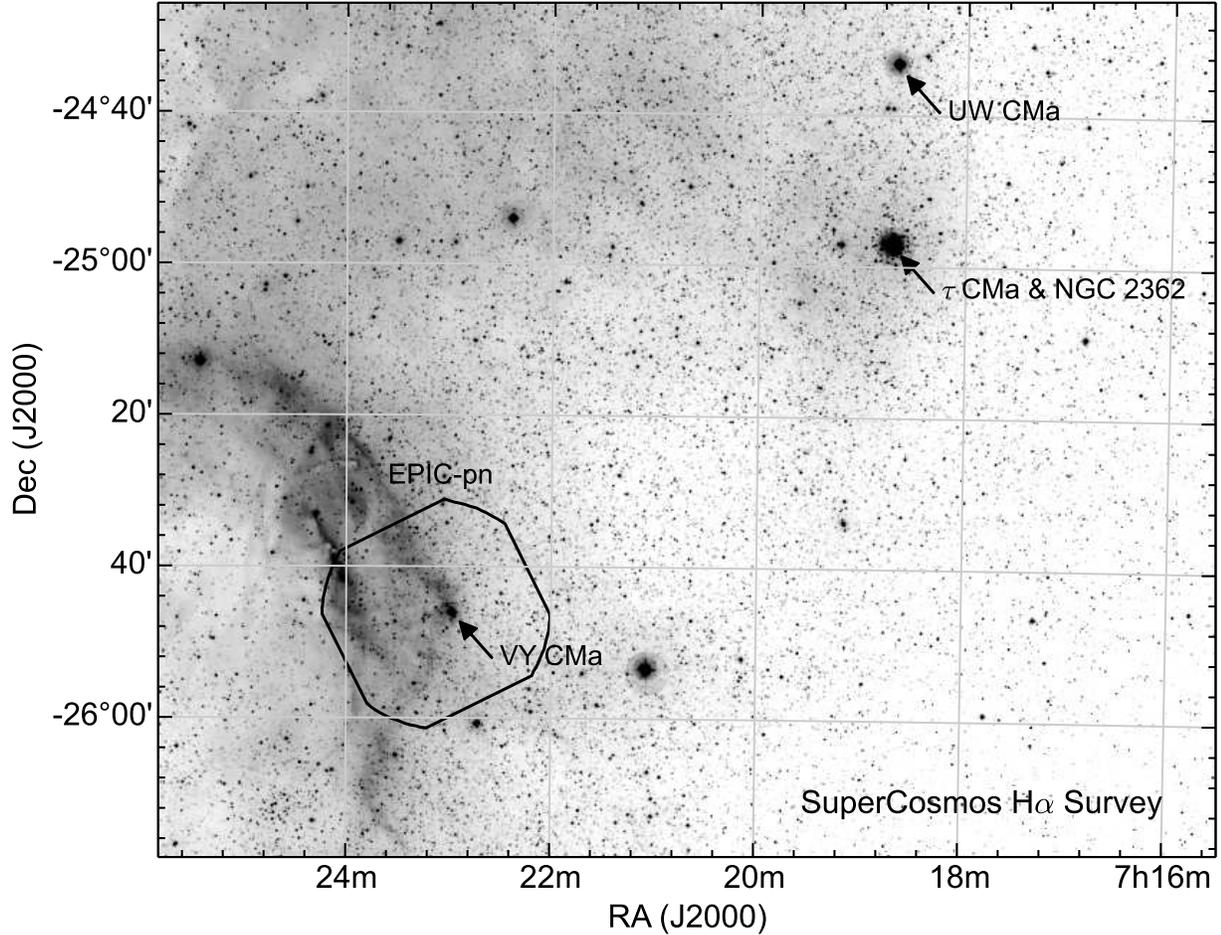}
\caption{H-$\alpha$ image mosaic of a $2.3\times1.8$ square degree field encompassing VY CMa, NGC 2362, and the semidetached eclipsing binary UW CMa, constructed from images acquired by the SuperCOSMOS H-$\alpha$ Survey \citep{2005MNRAS.362..689P}. The field of the XMM-Newton EPIC-pn observation analyzed in this paper is indicated. \label{fig1}}
\end{figure*}

To determine whether such strong magnetic fields might manifest themselves in the form of coronal X-ray activity, we used XMM to search for X-ray emission from VY CMa. 
These X-ray observations have the additional benefit that they can be used to assess whether the larger environment around VY CMa --- which includes a large molecular cloud and the young stellar cluster NGC 2362 \citep[Figure~\ref{fig1};][]{1978ApJ...219...95L} --- might be populated by late-type, pre-main sequence stars that formed during the same episode of star formation that  spawned the hypergiant itself. 
Such young, lower-mass stars should feature high levels of coronal activity and, hence, relatively large X-ray luminosities \citep[e.g.,][]{2006A&A...460..133D}. 

\section{Data and Analysis}

\subsection{Observations}

VY CMa was observed by the XMM-Newton satellite observatory on 07 May 2012 (ObsID: 0691190101).  
X-ray data were obtained with the European Photon Imaging Camera (EPIC), which is comprised of three X-ray imaging spectrometers  (pn, MOS1, MOS2), and the X-ray dispersion grating spectrometers (RGS). Optical/UV imaging was performed with the optical monitor (OM) through visible (V; $\lambda_{\rm eff} = 543 {\rm ~nm}$) and ultraviolet filters (UVW1; $\lambda_{\rm eff} = 291 {\rm ~nm}$).  
EPIC images were performed with the \verb+THICK+ optical blocking filter to ensure optical photons from VY CMa would not lead to a spurious detection. 
The EPIC detectors provide sensitivity in the 0.15 to 12 keV energy range with moderate energy resolution ($E/\Delta E\sim 20\rm{-}50$) at spatial resolution (point spread function core FWHM) of up to $\sim 6$-$12^{\prime\prime}$, albeit with broad half-encircled-energy widths up to $15^{\prime\prime}$. 
EPIC detectors pn, MOS1, and MOS2 were exposed for 20.0, 21.6, and 21.6 ks, respectively.
The RGS and the OM data are unusable for our purposes due to lack of source photon detections and artifacts caused by the optical brightness of VY CMa, respectively.
The apparent detection of VY CMa in the observation with the UVW1 filter is spurious and caused by residual charge from the saturated V filter observation.
Hence, we focus our analysis on the EPIC observations and, in particular, on the pn data, as the pn instrument has the highest sensitivity over the energy range of X-ray emission expected from stellar coronal activity (i.e., a thermal plasma at temperatures of a few $10^{6}$~K to a few $10^{7}$~K).

We used the XMM-Newton pipeline products for our analysis. 
These Processing Pipeline System (PPS) products include ``cleaned'' event lists for the EPIC instruments; lists of X-ray sources detected in the field (via the PPS automated source detection routine); and X-ray spectra and light curves for sources with at least 500 total-band EPIC counts, PSF-weighted detector coverage better than 50\%, and total-band detection likelihood $\geq15$ \citep{2009A&A...493..339W}. 
The PPS processing identified 20 X-ray emitting field sources in the vicinity (within $\sim5^{\prime}$) of VY CMa and 107 in the entire field of view. 
The PPS product source lists include count rates determined within soft (S), medium (M), and hard (H) bands defined over the energy ranges 0.5-1.0 keV, 1.0-2.0 keV, and 2.0-4.5 keV, respectively. 
Two hardness ratios are calculated from these three bands, HR1 = (M-S)/(M+S), and HR2 = (H-M)/(H+M), and are used in our analysis (\S\ref{fieldsource_section}). 

\subsection{Limits on the X-ray flux from VY CMa\label{vycmalimits}} 

We constructed images within various energy bands (0.5-8.0 keV, 0.5-1.0 keV, 1.0-2.0 keV, 2.0-4.0 keV) to limit background, but found no evidence for X-ray emission from VY CMa in any of these images.  
We also extracted source and background spectra to search for spectroscopic evidence of X-ray emission from VY CMa. 
Five source-free circular regions, each with a radius of  $20^{\prime\prime}$  and restricted to the pn CCD containing VY CMa (i.e., \verb+CCD_ID+=7), were selected to define the background. 
The background-subtracted spectrum does not reveal any signal above the background.  
We conclude that VY CMa is not detected in the EPIC observations.

To quantify the X-ray nondetection of VY CMa, we have estimated the source count rate required to produce a $2\sigma$ detection above the background in the EPIC-pn observation. 
The background count rate is determined by extracting the counts in the aforementioned background regions.  
Within the energy range 0.5-8.0 keV, we found a total of 197 background counts from these five regions, for a normalized background count rate of $1.57\times10^{-6}~{\rm counts~s}^{-1}~{\rm arcsecond}^{-2}$.  
Scaling this background rate to a source region at RA [J2000] = 07:22:58.33, Dec [J2000] = -25:46:03.24 with a radius of $30^{\prime\prime}$, we predict $89\pm9$ background counts at the position of VY CMa, consistent with the measured background of $91\pm9$ counts at this position. 
Hence, we confirm the lack of detection of VY CMa in the XMM observation.
The background-subtracted count rate required to produce a  $2\sigma$  detection in the 0.5 to 8 keV energy range is $1.13\times10^{-3}~{\rm counts~s}^{-1}$, after correcting for the encircled energy fraction of 80\% expected for the $30^{\prime\prime}$ extraction region on the EPIC-pn observation. 

\begin{figure}
\centering
\includegraphics[scale=0.5]{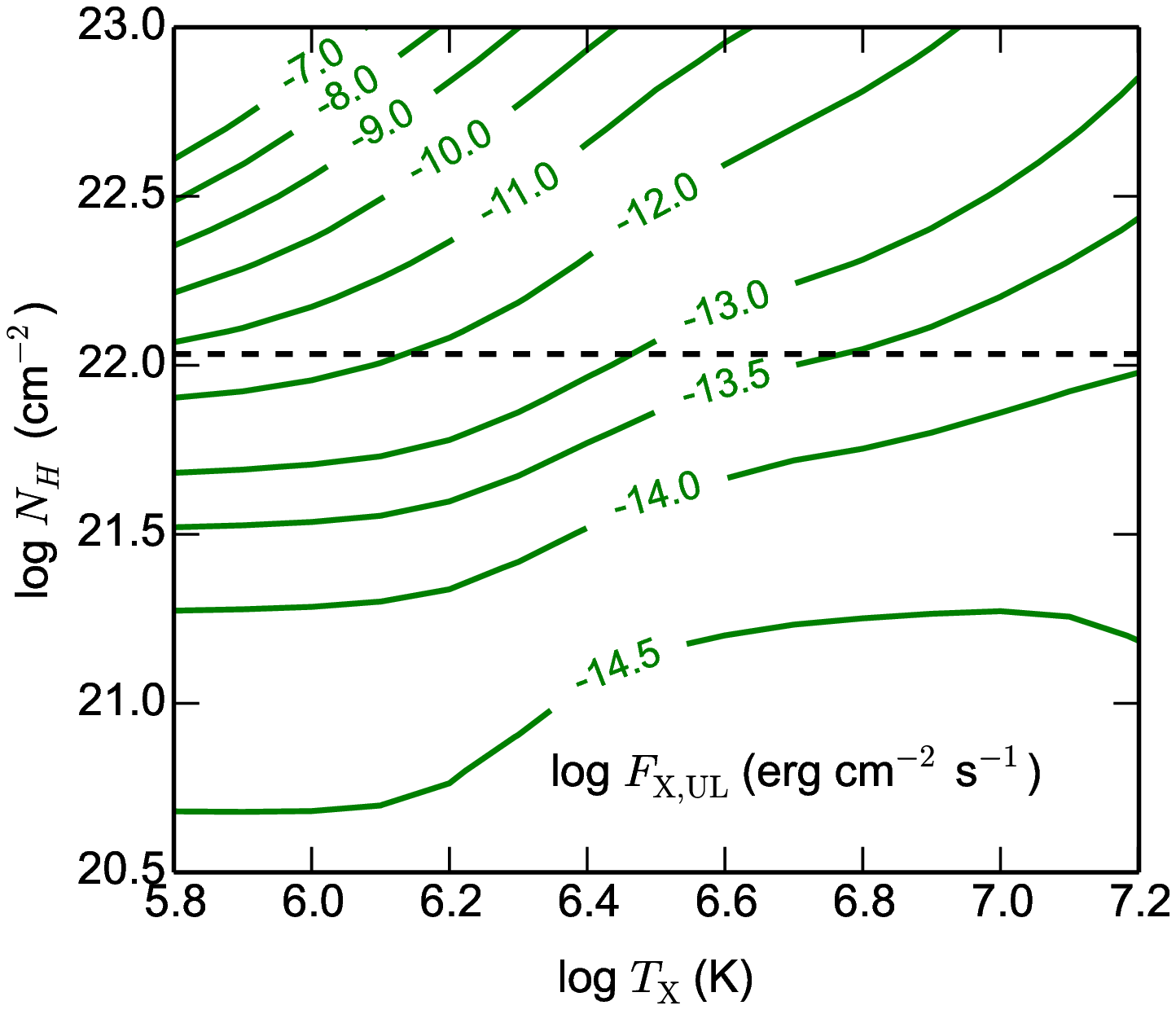}\\
\includegraphics[scale=0.5]{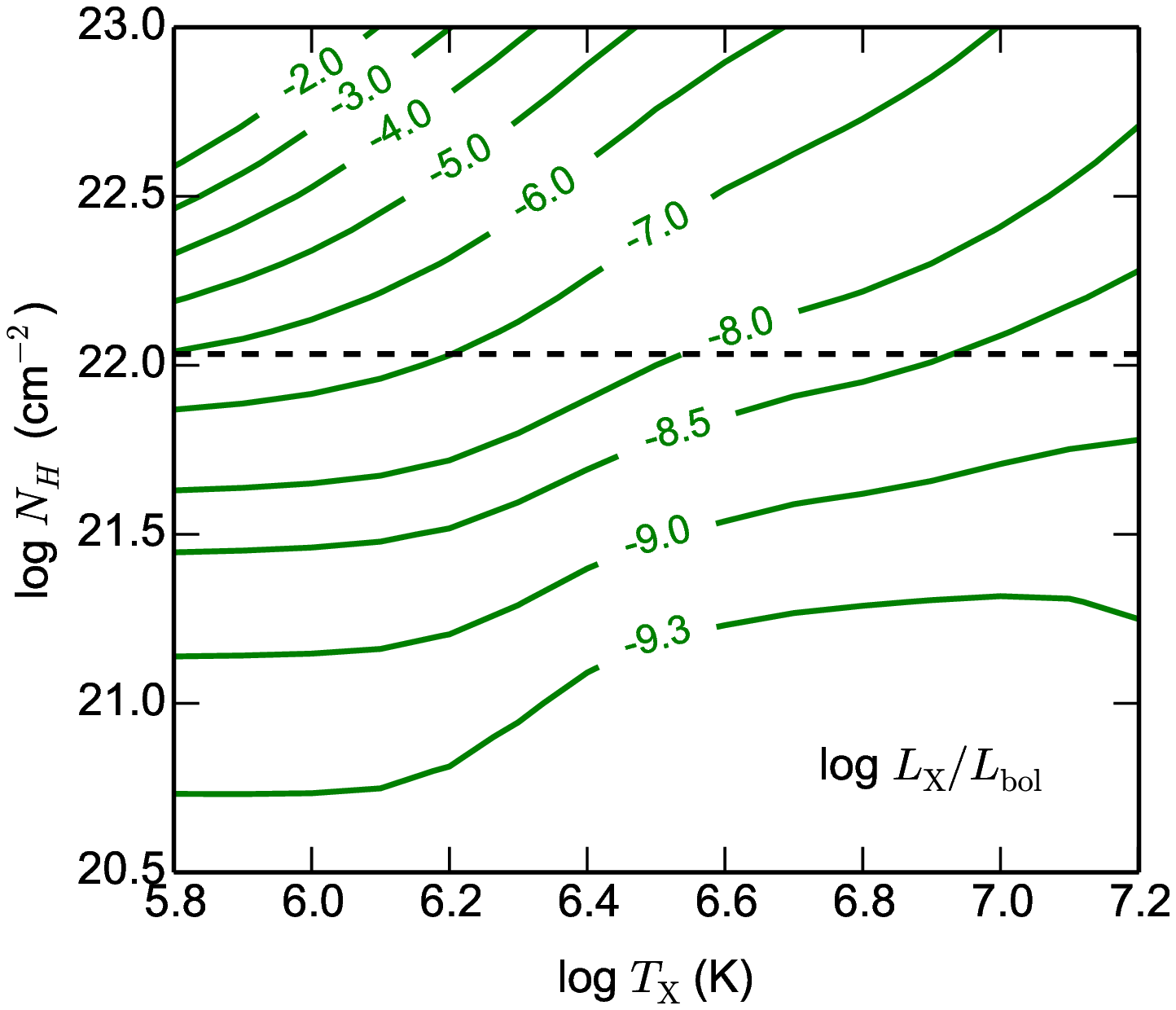} 
\caption{Upper limit calculations on the $N_H-T_X$ plane obtained from our XMM EPIC-pn observation of VY CMa. The contours indicate upper limits on source flux $F_{X,UL}$ (top) and relative X-ray luminosity $L_X/L_{\rm bol}$ (bottom). The vertical dashed lines in each panel indicates the total (interstellar and circumstellar) absorption based on the visual extinction \citep{2007AJ....133.2716H}. \label{figlxlbol}} 
\end{figure}

To determine the sensitivity to X-ray flux corresponding to this background-subtracted $2\sigma$ count rate, we must assume an underlying model. 
We hence adopt an absorbed, optically-thin thermal plasma \citep[MEKAL;][]{1985A&AS...62..197M, 1986A&AS...65..511M, 1992Kaastra, 1995ApJ...438L.115L} that effectively describes a variety of physical processes, including coronal activity and shocks in a stellar wind. 
For generality, we have computed X-ray fluxes for a grid of absorbing columns, $N_H$, and plasma temperatures, $T_{\rm X}$, that covers a wide range of values ($\log N_H [{\rm cm}^{-2}] = 20 {\rm ~to~} 23$ and $\log T_{\rm X} [{\rm K}]$ from 5.75 to 7.25). 
Lower temperatures are typical of solar--like coronal activity while the higher temperatures are typical of more energetic flare-like activity associated with coronal mass ejections \citep{2004A&A...424..677O}. 
Additionally,  X-ray emission from wind shocks associated with early-type massive stars is typically characterized by plasma temperatures of a few $10^{6}$~K and lower ISM absorbing columns \citep{2011ApJS..194....7N}. 
We perform PIMMS simulations over the entire $N_H-T_{\rm X}$ grid to estimate the $2\sigma$ upper limit on X-ray flux, $F_{\rm X, UL}$, from the $2\sigma$ background-subtracted count rate.
Assuming a distance of 1.3 kpc and $L_{\rm bol}\sim 3\times10^{5}~L_{\odot}$, we then calculated upper limits for $L_{\rm X}/L_{\rm bol}$, a quantity often used as a stellar magnetic activity indicator. 
The results of these simulations are presented in Figure~\ref{figlxlbol}.

The total interstellar and circumstellar visual extinction, $A_V$, to VY CMa is estimated by \citet{2007AJ....133.2716H} as 5-6 mag. This suggests an intervening absorbing column of $\sim$$10^{22} {\rm ~cm}^{-2}$. 
For such a column, at a nominal $T_X$ of $3\times10^{6}$~K, we estimate an unabsorbed $F_{\rm X, UL} = 8 \times 10^{-14} {\rm ~erg~cm}^{-2} {\rm ~s}^{-1}$, $L_{\rm X} <  1.6 \times 10^{31} {\rm ~erg~s}^{-1}$, and $\log L_{\rm X}/L_{\rm bol} < -8$.

\subsection{Classification of the X-ray Sources within the XMM Observation Field of View\label{fieldsource_section}}

We have cross-referenced all 107 X-ray sources detected in our XMM observation of VY CMa with infrared and astrometric catalogs: Two-Micron All Sky Survey Point Source Catalog \citep[2MASS;][]{2006AJ....131.1163S}, Wide-field Infrared Survey Explorer source catalog \citep[WISE;][]{2010AJ....140.1868W}, and the United States Naval Observatory CCD Astrograph Catalog \citep[UCAC4;][]{2013AJ....145...44Z}. 
The positional uncertainties of most these X-ray sources, according to the PPS output, is less than $\sim3^{\prime\prime}$. 
Hence, we have identified potential IR counterparts for the PPS-identified X-ray sources using a correlation radius of 6$^{\prime\prime}$ and selecting the closest match to the position of the X-ray source. 
Within this correlation radius, we find 47 X-ray sources have potential IR counterparts.
Ultimately, we dismissed the WISE point source data because VY CMa saturated the field with stray light. This stray light contamination is especially problematic at the longest WISE wavelengths.

\begin{figure}
\centering
\includegraphics[scale=0.5]{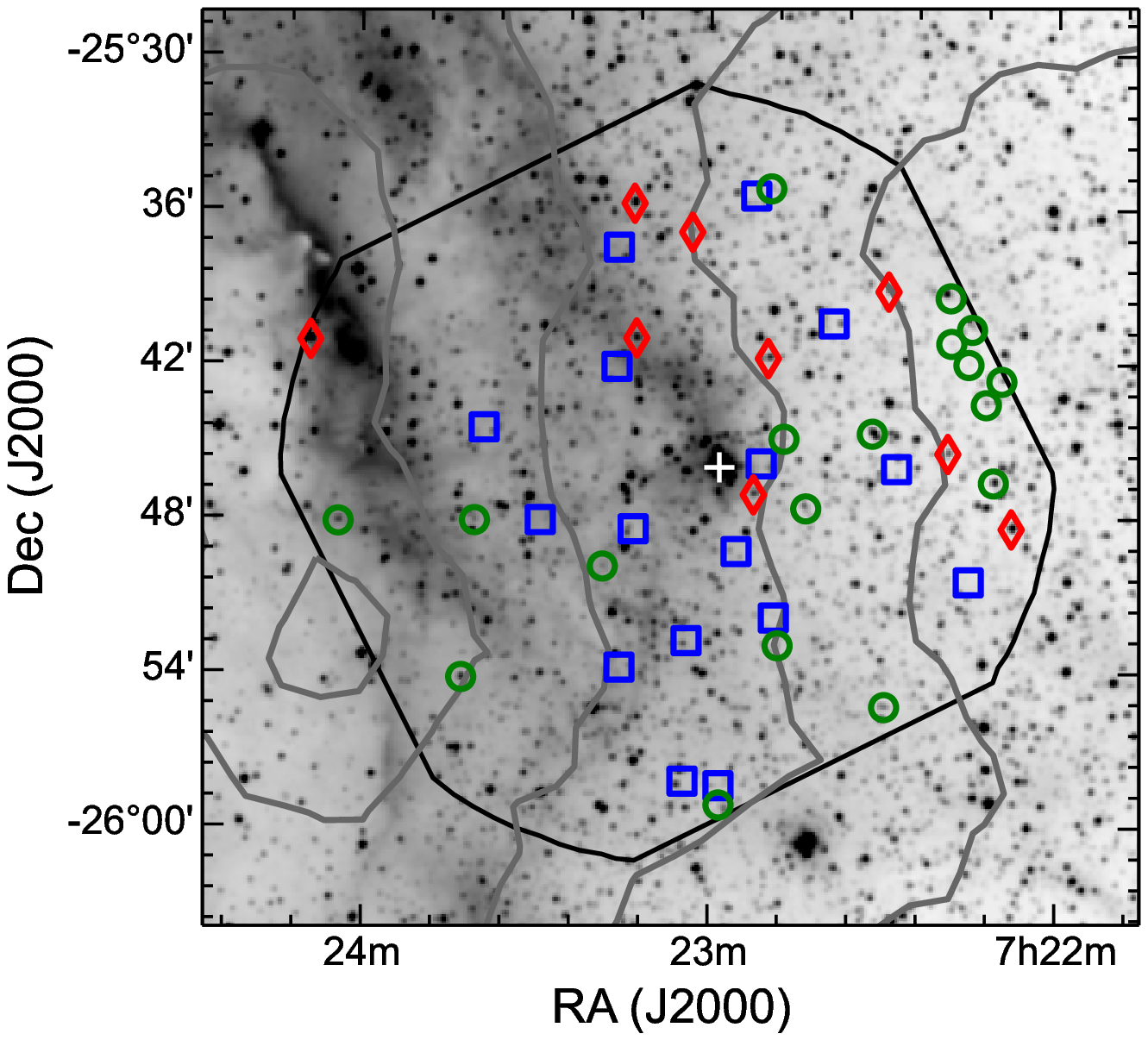}\\
\includegraphics[scale=0.5]{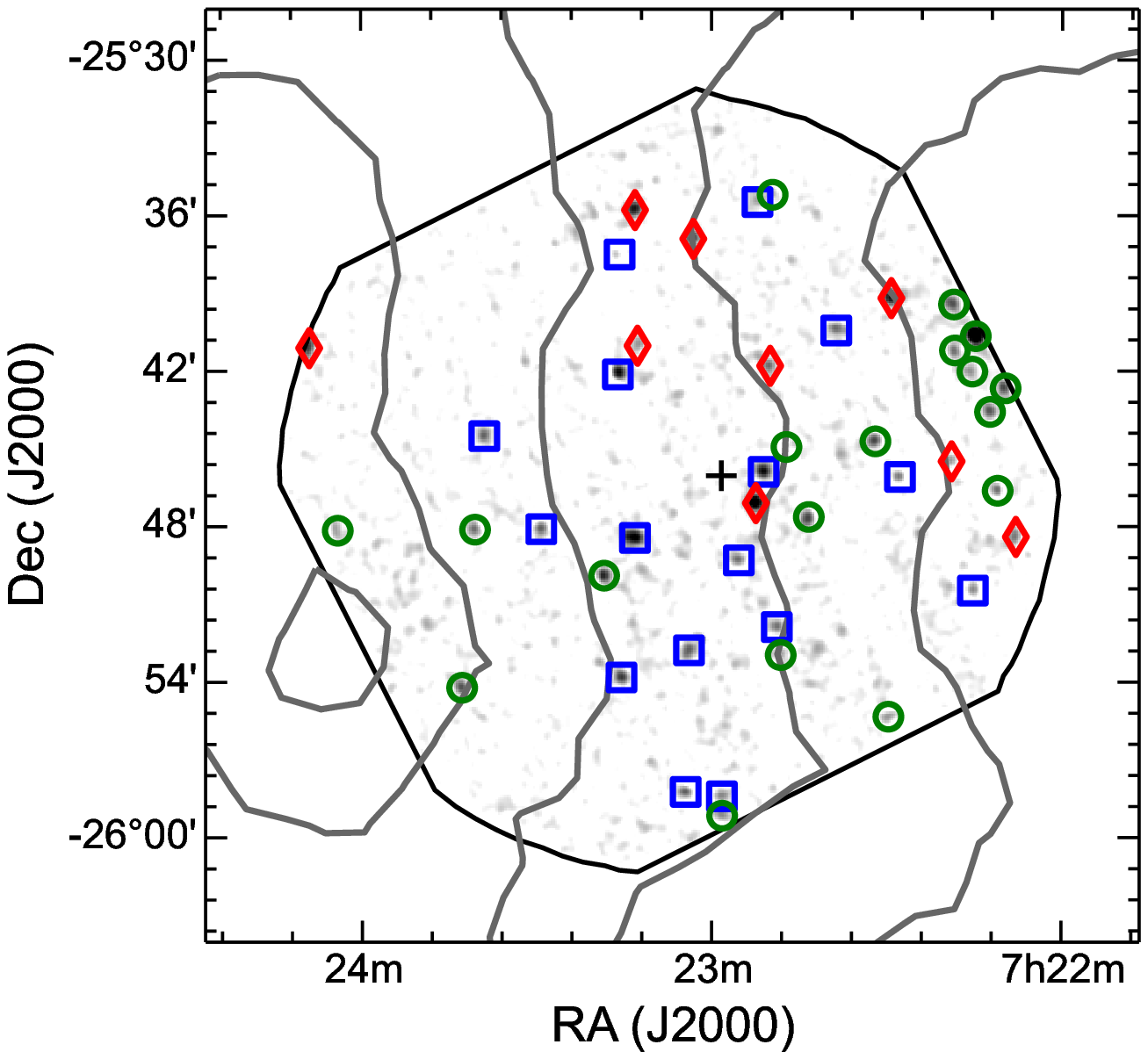}
\caption{SuperCOSMOS H-$\alpha$ image (top) and XMM-Newton EPIC pn X-ray 0.5-8.0 keV image (bottom) of the VY CMa field. In each image, we have indicated the position of VY CMa by a cross and X-ray sources from the XMM observations by open symbols. Green circles indicate X-ray sources with IR counterparts, red diamonds are X-ray sources with significant proper motions, and blue squares are X-ray sources without IR counterparts. Contours corresponding to $E_{\rm B-V} = $ 0.5, 0.6, 1.0, 2.0, and 3.0 mag from the galactic extinction map of \citet{1998ApJ...500..525S} are represented by grey contours with extinction generally increasing from right to left. \label{fig2}}
\end{figure}

We examined the total band detection likelihood (\verb+PN_DET_ML+) as determined by the  PPS source-detection algorithm to filter the source list to the most significant detections.  
We required that \verb+PN_DET_ML+$\geq 20$, resulting in a sample of 43 well-detected X-ray sources, 27 of which have potential IR counterparts. 
The subsamples of these 43 X-ray sources with and without IR counterparts are illustrated in Figure~\ref{fig2} and are described in Tables~\ref{tblhra}~\&~\ref{tblhrb}, respectively.

At the distance to VY CMa (1.3 kpc), any source with a reliable measurement of proper motion is unlikely to be associated with the hypergiant. 
We identify nine IR counterparts among the 27 X-ray sources with potential IR counterparts that display evidence for significant (measurable) proper motions. 
Specifically, for each of these nine sources, the magnitude of proper motion error is less than 68\% of the magnitude of the proper motion measurement itself. 
We do not consider these nine objects with significant proper motions further in this paper. 

\section{Discussion}

\subsection{Constraints on the magnetic activity of VY CMa} 

\begin{figure}
\centering
\includegraphics[scale=0.55]{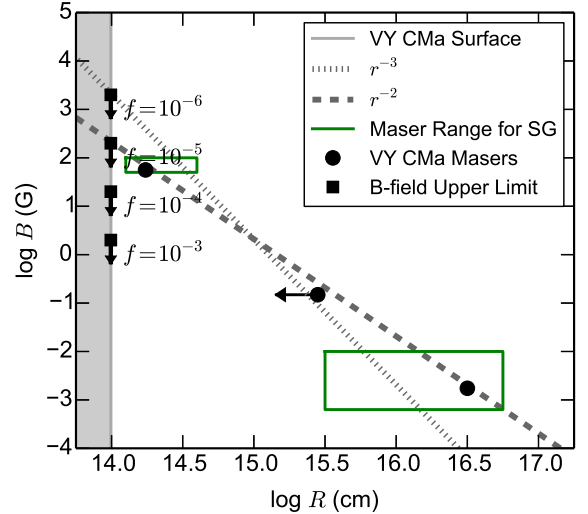}
\caption{The limits on the surface magnetic field strength of VY CMa as determined from our XMM X-ray non-detection. 
The surface magnetic field strength upper limits determined from our non-detection of X-ray emission, $\log L_X/L_{\rm bol} = -8$ found by assuming $\log N_H ({\rm ~cm}^{-2}) = 22$ and $\log T_X (K) = 6.5$ are overlaid for a few filling factors, $f$.  
We show magnetic field strengths inferred from maser measurements \citep[see][]{2002A&A...394..589V}, the radial behavior for a solar-like ($r^{-2}$) and dipole-like ($r^{-3}$) magnetic field.  The ranges of magnetic fields from maser measurements of all supergiants are shown by the green boxes. \label{figbfield}} 
\end{figure}

Circular polarization measurements of masers around mass-losing AGB stars and supergiants provide estimates of magnetic fields at various displacements from the stars that are determined by the formation and survival of the observed molecules \citep{2002A&A...394..589V,2005MmSAI..76..462V}. 
In many of these evolved star envelopes, the ensemble of masers at various radii from the star $r$ suggests that the magnetic field strength $B$ drops off as $r^{-2}$ or $r^{-3}$, indicative of a solar-like or dipole-like magnetic field, respectively. 
The circular polarization measurements of VY CMa indicate that the magnetic field configuration may follow a solar-like magnetic field topology (see Figure~\ref{figbfield}), which suggests the surface magnetic field strength could be $\sim$200~G. 
The total surface magnetic flux, $\Phi_{\rm M}$, is given by $\Phi_M = 4 \pi R^2 f B$, where the filling factor, $f$, parameterizes how $\Phi_{\rm M}$ is diluted by the relative size of the active areas on the surface. 
For a stellar radius appropriate for VY CMa and an arbitrary filling factor, the expected surface magnetic flux is then 
\begin{equation}
\Phi_{\rm M} = 2\times10^{26} {\rm ~Mx} \left( \frac{f}{10^{-5}} \right) \left( \frac{B}{200 {\rm ~G~}} \right) \left( \frac{R}{1400 R_{\odot}} \right)^{2}, 
\end{equation}
where Mx is the maxwell unit ($1 {\rm ~Mx} = 1 {\rm ~G~cm}^{2}$). 

To compare this estimate to our upper limit on the X-ray luminosity, we use an empirical relationship between stellar X-ray luminosity, $L_X$ and magnetic flux, $\Phi_{\rm M}$, that is found to hold over several orders of magnitude \citep{2003ApJ...598.1387P}. We caution that this relationship describes the behavior of active regions on the Sun and late-type stars and, hence, it  is not clear if the same relationship would apply to magnetically-driven coronal activity on a supergiant. 
\citet{2013ApJ...779..183F} provide a recent update of this relationship that  is constrained by the largest number of low-mass stellar luminosity and magnetic field measurements to date; specifically, 
\begin{equation}
\log \Phi_{\rm M} = (11.86\pm0.68) + (0.459\pm0.018) \log L_{\rm X}.
\end{equation}
With their formulation, our upper limit on $L_{\rm X}$ of $2\times10^{31} {\rm ~erg~s}^{-1}$ --- which is based on Figure~\ref{figlxlbol} for $\log T_{\rm X} (K) = 6.5$ and $\log N_H ({\rm cm}^{-2}) = 22$ --- suggests $\Phi_{\rm M} \lesssim 2\times10^{26} {\rm ~Mx}$. 
We remark that for $\log N_H ({\rm cm}^{-2}) > 22$ and $\log T_X (K) = 6.5$, the upper limit on $L_{\rm X}$ becomes proportional to $N_H^{2.8}$, whereas $\Phi_{\rm M} \propto N_H^{1.4}$; hence, where inferred upper limits are concerned, the value for $\Phi_{\rm M}$ is less sensitive than $L_{\rm X}$ to the value adopted for $N_H$. 

Overall, based on these considerations, the upper limit on the X-ray luminosity suggests that $fB \lesssim 2\times10^{-3} {\rm ~G}$ for VY CMa. 
In Figure~\ref{figbfield} we illustrate the surface magnetic field strength upper limits for various choices of filling factor $f$.
Note that, adopting the maser-based estimate for the surface magnetic field, our X-ray upper limit indicates that $f\lesssim 10^{-5}$. 
Such an active region size is still substantial, given the enormous surface area of VY CMa; indeed, this filling factor upper limit corresponds to an active region comparable in size to the Earth-facing surface of a star with a radius of $4 R_{\odot}$.

In the absence of constraints on the sizes of potential active regions on VY CMa, it is impossible to break the foregoing degeneracy between $f$ and $B$. 
We note, however, that the presence of active regions is suggested by interferometric measurements that yield evidence for bright spots on the disks of a few red supergiants \citep{1997MNRAS.285..529T}. 
In these objects, the bright spots have likely filling factors of a few tenths, which would suggest a low surface field strength assuming VY CMa is characterized by the same regime of $f$. 
On the other hand, the episodic nature of the mass loss events from VY CMa is suggestive of periods of intense magnetic activity that generate eruptions. These eruptive events could be analogous to coronal mass ejections during periods of high solar activity, but with typical active-period timescales of decades rather than years \citep[e.g.,][]{2014arXiv1410.1622O}. Our X-ray nondetection could then indicate that VY CMa is not presently in such a magnetically active state. 

\subsection{X-ray Observations of Hypergiants} 

The yellow hypergiant IRC+10420 is the only other hypergiant that has been targeted thus far by a modern X-ray observatory. 
The non-detection of X-ray emission from IRC+10420 based on XMM-Newton observations was reported in \citet{2014NewA...29...75D}. 
However, these authors report upper limits on the {\it observed} X-ray flux ($\lesssim 1-3 \times 10^{-14} {\rm ~erg~cm}^{-2} {\rm ~s}^{-1}$); as a result, their inferred relative source X-ray flux ($f_{\rm X}/f_{\rm bol} \lesssim 1.1-3.5\times10^{-8}$) does not take into account the affects of intevening absorption. 
The large distance to IRC+10420 \citep[$D\sim5{\rm ~kpc}$][]{1993ApJ...411..323J} results in large interstellar absorption ($A_V\sim6{\rm ~mag}$), which dominates the total absorption to the star; whereas, for VY CMa, the circumstellar absorption likely dominates \citep{2007AJ....133.2716H}.  
The circumstellar absorption toward IRC+10420 provides an additional $\sim 1 {\rm ~mag}$ of absorption ---  this modest extinction likely reflects our nearly pole-on view of the predominantly equatorial circumstellar material \citep{2010AJ....140..339T}.
Accounting for the likely absorbing column toward IRC +10420 (at  least $N_H\sim 10^{22} {\rm ~cm}^{-2}$ of interstellar and circumstellar absorption) and assuming a plasma temperature appropriate for coronal and self-shocking winds ($T_{\rm X} \sim 3\times10^{6} {\rm ~K}$), PIMMS simulations give the upper limit on the unabsorbed intrinsic X-ray flux from IRC+10420 as  $F_{\rm X, UL} \lesssim 10^{-12} {\rm ~erg~cm}^{-2}{\rm ~s}^{-1}$. 
The upper limit on the relative X-ray luminosity then becomes $L_{\rm X}/L_{\rm bol} \lesssim 10^{-6}$. 
These rather modest constraints, relative to those we have obtained for VY CMa ($F_{\rm X, UL} \lesssim 8\times10^{-14} {\rm ~erg~cm}^{-2}{\rm ~s}^{-1}$ and $L_{\rm X}/L_{\rm bol} \lesssim 10^{-8}$, respectively; \S\ref{vycmalimits}), reflect in large part the low sensitivity of the XMM observations of IRC +10420 \citep[this XMM exposure was only $\sim$10 ks in duration and was compromised by high particle backgrounds;][]{2014NewA...29...75D}.

Next, we compare the relative X-ray luminosities of VY CMa and IRC+10420 with those of the self-shocking winds of OB stars and W-R stars. 
Self-shocking winds of OB stars and W-R stars give rise to X-ray emission within the wind and typically feature plasmas of a few MK with $L_{\rm X}/L_{\rm bol} \sim 10^{-7}$ \citep[e.g.,][]{2011ApJS..194....7N}. 
These measurements are corrected for absorption by the ISM but not for additional self-absorption by the wind.
Indeed, there is theoretical evidence that thin-shell mixing in O-star radiative wind shocks  can lead to decreased $L_{\rm X}/L_{\rm bol}$ ratios, especially in denser winds of earliest O types and W-R star winds \citep{2013MNRAS.429.3379O}. 
Hence, caution is advised when making a comparison between OB stars and the red and yellow hypergiants considered here. 
Adopting the standard treatment of $L_{\rm X}$ measurements of OB stars and W-R stars, we consider the upper limit resulting from only interstellar extinction.
For IRC+10420, most of the absorption arises from the ISM and $L_{\rm X}/L_{\rm bol}\lesssim10^{-6}$, whereas for VY CMa, the circumstellar ejecta dominates the absorbing column to the star, so according to Figure~\ref{figlxlbol} and an ISM column of $\log N_H ~({\rm cm}^{-2})\sim21.25$, $L_{\rm X}/L_{\rm bol} \lesssim 10^{-9}$.
Such a low ratio suggests that VY CMa does not exhibit energetic (X-ray emitting) wind shocks in its ejecta. 
This result is not particularly surprising, given that its envelope terminal velocity is $\sim$35 km s$^{-1}$ \citep{2005AJ....129..492H}, i.e., more than an order of magnitude lower than the typical wind speeds of hot, massive stars (whose wind speeds of hundreds to thousands of km s$^{-1}$ are compatible with X-ray-producing shocks).
We caution that our observation of VY CMa is not sensitive to lower-speed (hence lower-temperature) shocks, which are more susceptible to absorption by circumstellar material (see Figure~\ref{figlxlbol}). 
On the other hand, after accounting for the ISM absorbing column toward IRC+10420, the XMM observations reported by \citet{2014NewA...29...75D} cannot constrain whether energetic shocks are present in the winds from IRC+10420 \citep[whose envelope terminal velocity is $\sim$70 km s$^{-1}$,][]{2009A&A...507..301D}. 

\begin{figure}
\centering
\includegraphics[scale=0.5]{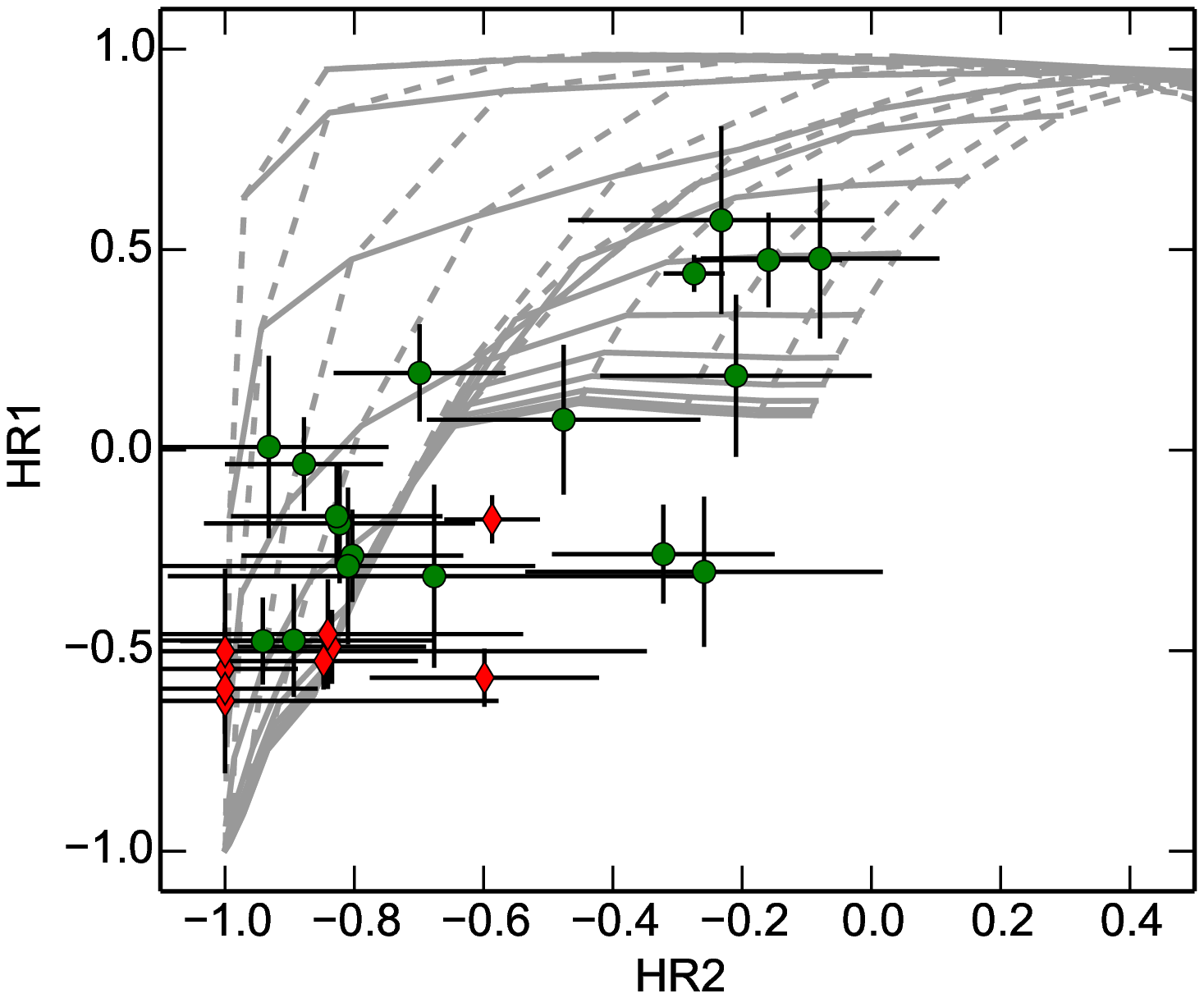}\\
\includegraphics[scale=0.5]{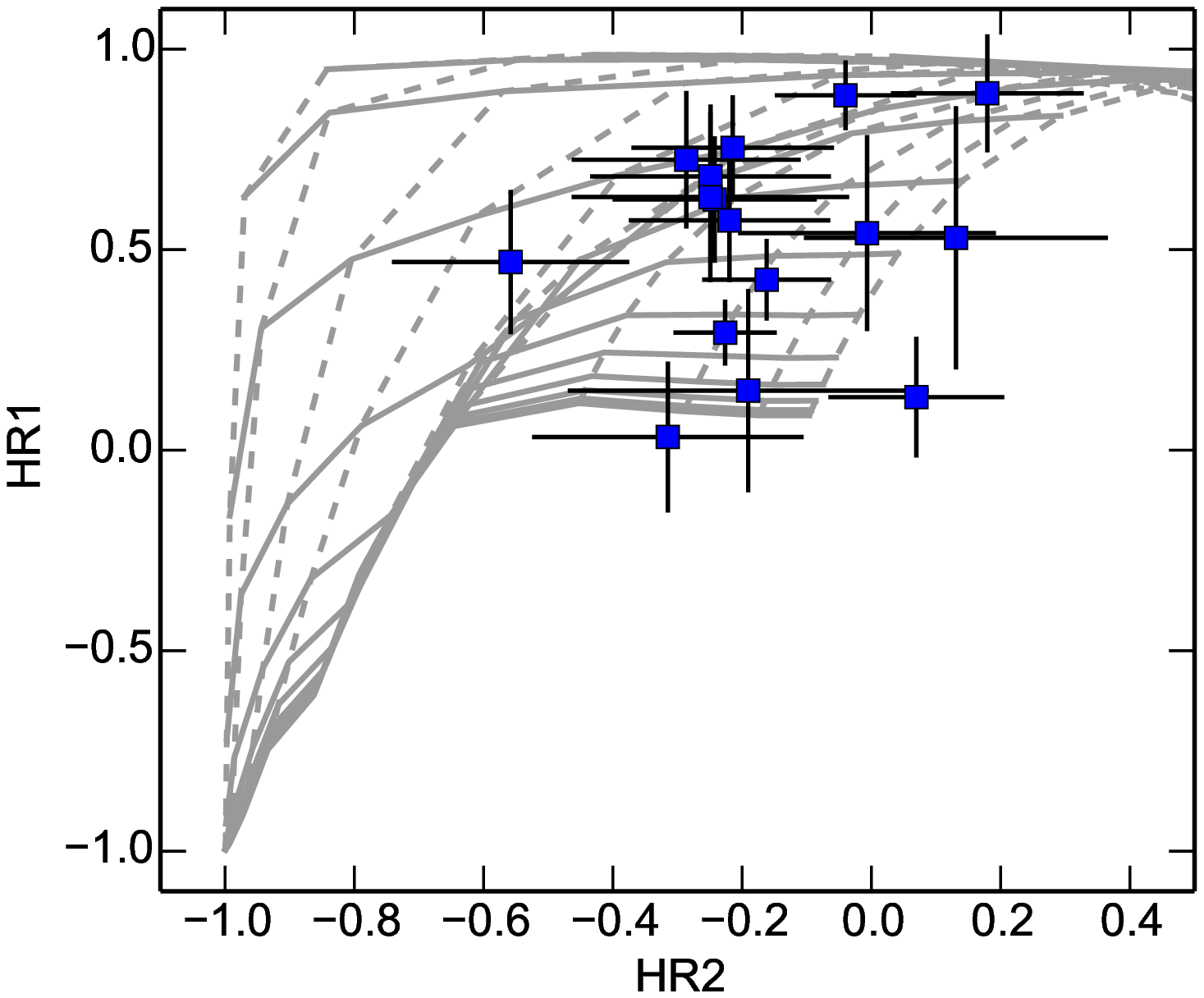}
\caption{Hardness ratios of the strongest X-ray sources within $15^{\prime}$ of VY CMa. Symbols are as in Figure~\ref{fig2}. X-ray emitting sources with IR counterparts are shown in the upper panel and those without IR counterparts are shown in the lower panel. A grid of absorbed, optically-thin thermal plasma models are overlaid as gray lines.
The column densities from $10^{20}$ to $3\times10^{23} {\rm ~cm}^{-2}$ (solid lines) increase upwards while the plasma temperatures $10^{6}$ to $10^{8}$~K (dashed lines) increase  from left to right. \label{fighr}} 
\end{figure}

\begin{figure}
\centering
\includegraphics[scale=.5]{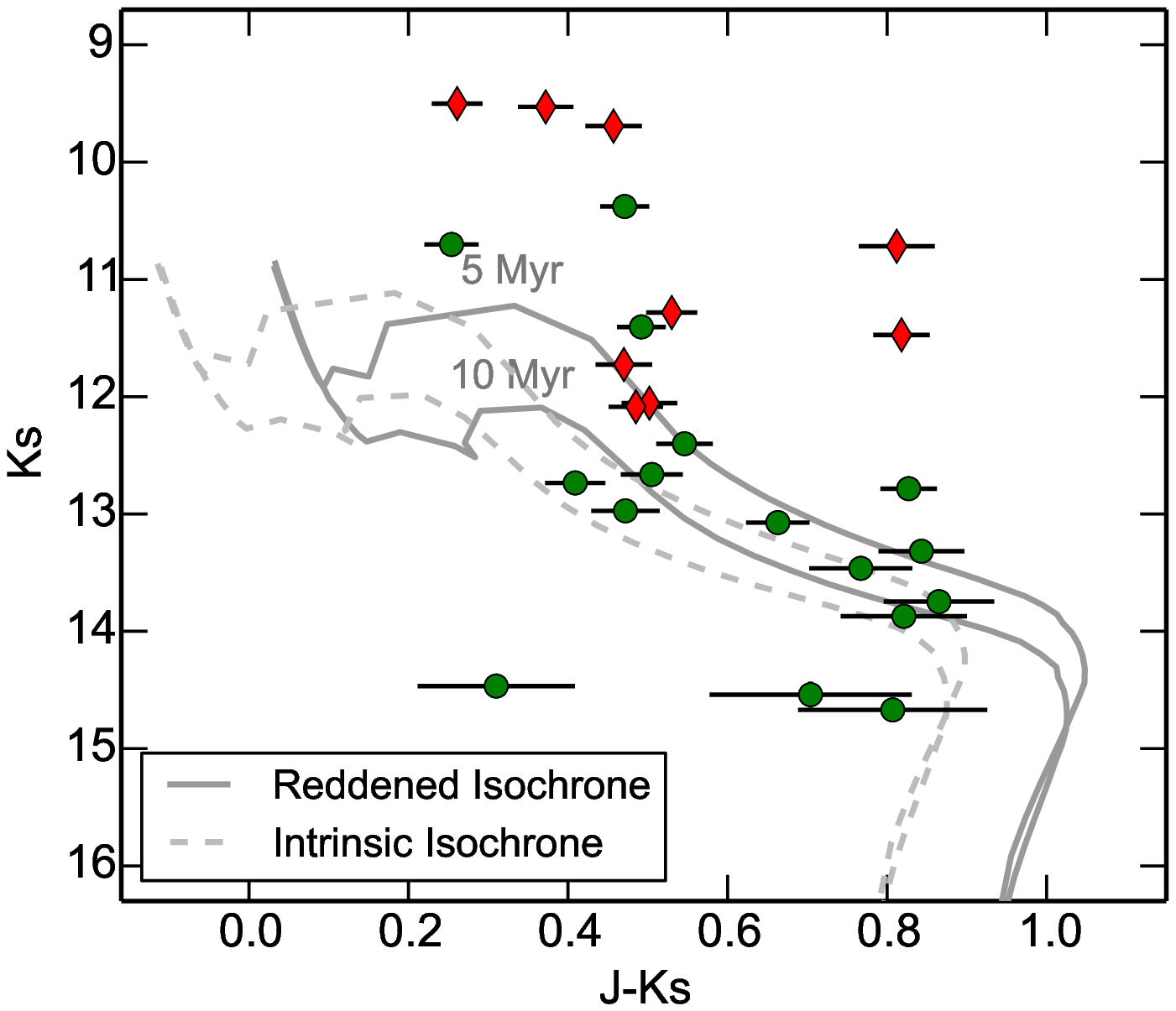}\\
\includegraphics[scale=.5]{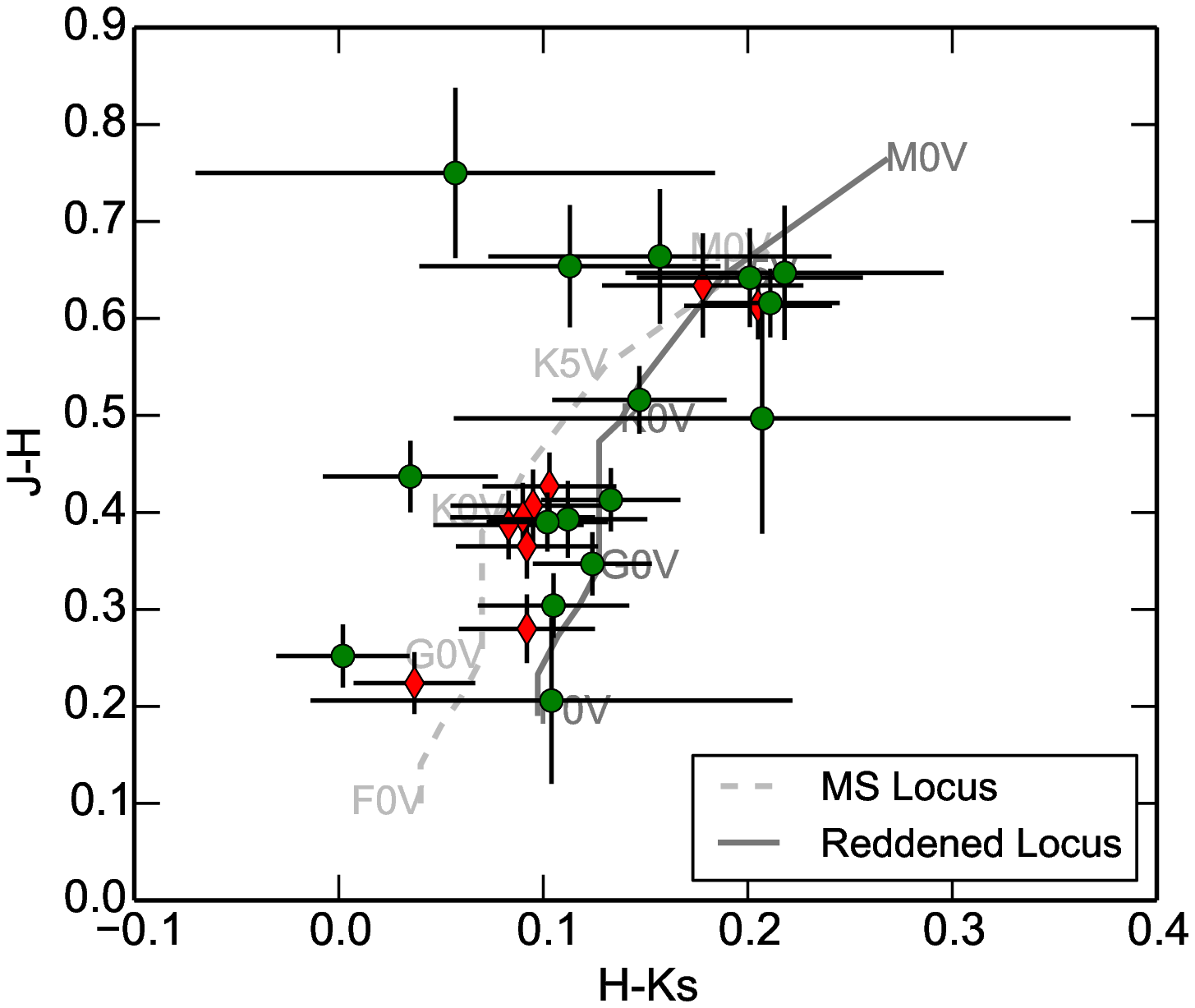}
\caption{Observed 2MASS color-magnitude (top) and color-color (bottom) diagrams of the X-ray sources with IR counterparts. Pre-main sequence isochrones for ages of 5 and 10 Myr (lower), scaled to the distance of VY CMa, are overplotted on the color-magnitude diagram; solid and dashed lines indicate isochrones with and without reddening of $A_V \sim 1.5$~mag (see text), respectively. 
The intrinsic main sequence (MS) locus \citep{2011MNRAS.418.2219S}, and the same locus reddened assuming $A_V \sim 1.5$~mag, are overplotted on the color-color diagram.  Symbols for IR counterparts to X-ray sources are as in Figure~\ref{fighr}.\label{figcmd}} 
\end{figure}

\begin{deluxetable*}{lccccccccccccccc}
\tabletypesize{\scriptsize}
\tablecaption{X-ray Sources with IR counterparts\label{tblhra}} 
\tablewidth{0pt}
\tablehead{
\colhead{ID} & \colhead{RAJ2000} & \colhead{DECJ2000} & \colhead{$\theta_{*}^{a}$} & \colhead{log~$F_{\rm X}$} &
\colhead{HR1} & \colhead{HR2} & \colhead{$\theta_{\rm IR}^{b}$} & \colhead{K} & \colhead{J-H} &  \colhead{H-K} \\
 & \multicolumn{2}{c}{(decimal degrees)} & ($^\prime$) & (${\rm erg~cm}^{-2}{\rm ~s}^{-1}$) & & & ($^{\prime\prime}$) & (mag) & (mag) & (mag)
}
\startdata
1 & 110.695236 & -25.749485 & 2.8 & -13.36 & 0.19 & -0.21 & 4.8 & 14.47 & 0.21 & 0.10 \\
2 & 110.681101 & -25.795293 & 3.7 & -13.58 & -0.26 & -0.32 & 4.6 & 13.32 & 0.64 & 0.20 \\
3 & 110.827017 & -25.832432 & 6.0 & -13.14 & 0.47 & -0.16 & 2.3 & 14.54 & 0.50 & 0.21 \\
4 & 110.633655 & -25.745552 & 6.1 & -13.71 & -0.48 & -0.94 & 1.9 & 10.70 & 0.25 & 0.00 \\
5 & 110.701152 & -25.882938 & 7.3 & -14.32 & 0.01 & -0.93 & 1.7 & 13.87 & 0.66 & 0.16 \\
6 & 110.919803 & -25.802366 & 9.8 & -13.44 & -0.18 & -0.82 & 2.8 & 13.07 & 0.52 & 0.15 \\
7 & 110.576459 & -25.688474 & 10.2 & -13.42 & 0.48 & -0.08 & 5.6 & 14.90 & 0.15 & 1.17 \\
8 & 110.565065 & -25.701433 & 10.4 & -13.58 & 0.57 & -0.23 & 4.9 & 14.67 & 0.75 & 0.06 \\
9 & 110.551432 & -25.726772 & 10.6 & -13.67 & -0.16 & -0.83 & 1.3 & 12.66 & 0.39 & 0.11 \\
10 & 110.546044 & -25.777196 & 10.7 & -13.82 & -0.47 & -0.89 & 0.9 & 10.38 & 0.35 & 0.12 \\
11 & 110.706837 & -25.586718 & 11.0 & -14.14 & -0.31 & -0.68 & 1.2 & 13.46 & 0.65 & 0.11 \\
12 & 110.577615 & -25.657208 & 11.1 & -13.47 & -0.26 & -0.80 & 1.7 & 12.78 & 0.62 & 0.21 \\
13 & 110.561816 & -25.677843 & 11.2 & -12.34 & 0.44 & -0.27 & 0.9 & 14.37 & 0.86 & 0.82 \\
14 & 110.624613 & -25.923327 & 11.3 & -14.07 & -0.29 & -0.81 & 4.3 & 12.97 & 0.44 & 0.03 \\
15 & 110.540498 & -25.711559 & 11.5 & -13.49 & -0.03 & -0.88 & 1.5 & 12.40 & 0.41 & 0.13 \\
16 & 110.929428 & -25.904013 & 13.0 & -13.16 & 0.19 & -0.70 & 3.7 & 11.40 & 0.39 & 0.10 \\
17 & 110.743384 & -25.986525 & 13.1 & -13.74 & -0.30 & -0.26 & 3.0 & 13.75 & 0.65 & 0.22 \\
18 & 111.018006 & -25.802717 & 15.0 & -13.59 & 0.08 & -0.48 & 3.3 & 12.73 & 0.30 & 0.11 \\

\sidehead{Sources with Measurable Proper Motion:} 

1 & 110.719126 & -25.785464 & 1.7 & -13.20 & -0.17 & -0.59 & 2.5 & 12.05 & 0.41 & 0.10 \\
2 & 110.708815 & -25.696935 & 4.6 & -13.92 & -0.55 & -1.00 & 2.0 & 12.09 & 0.39 & 0.09 \\
3 & 110.803036 & -25.683911 & 6.0 & -13.98 & -0.50 & -1.00 & 0.6 & 11.28 & 0.43 & 0.10 \\
4 & 110.579420 & -25.757902 & 8.9 & -14.25 & -0.63 & -1.00 & 2.6 & 11.73 & 0.39 & 0.08 \\
5 & 110.763456 & -25.615153 & 9.2 & -13.49 & -0.49 & -0.83 & 1.3 & 11.47 & 0.61 & 0.21 \\
6 & 110.621950 & -25.653632 & 9.5 & -13.38 & -0.53 & -0.85 & 1.3 & 10.72 & 0.63 & 0.18 \\
7 & 110.805101 & -25.596622 & 10.8 & -13.16 & -0.57 & -0.60 & 1.6 & 9.69 & 0.37 & 0.09 \\
8 & 110.533127 & -25.807030 & 11.6 & -13.83 & -0.46 & -0.84 & 1.6 & 9.53 & 0.28 & 0.09 \\
9 & 111.038003 & -25.685400 & 16.7 & -13.26 & -0.59 & -1.00 & 3.0 & 9.50 & 0.22 & 0.04 
\enddata
\tablecomments{All measurements are observed and not dereddened.}
\tablenotetext{a}{The offset angle between the X-ray source and the position of VY CMa. }
\tablenotetext{b}{The offset angle between the X-ray source and the 2MASS source.}
\end{deluxetable*}

\begin{deluxetable*}{lccccccccccccccc}
\tabletypesize{\scriptsize}
\tablecaption{X-ray Sources without IR counterparts \label{tblhrb}} 
\tablewidth{0pt}
\tablehead{
\colhead{ID} & \colhead{RAJ2000} & \colhead{DECJ2000} & \colhead{$\theta_{*}^{a}$} & \colhead{log~$F_{\rm X}$} &
\colhead{HR1} & \colhead{HR2} & \colhead{$\theta_{\rm IR}^{b}$}  \\
 & \multicolumn{2}{c}{(decimal degrees)} & ($^\prime$) & (${\rm erg~cm}^{-2}{\rm ~s}^{-1}$) & & & ($^{\prime\prime}$) 
 }
\startdata
1 & 110.712895 & -25.765103 & 1.6 & -13.17 & 0.43 & -0.16 & 16 \\
2 & 110.730628 & -25.822038 & 3.3 & -13.54 & 0.72 & -0.29 & 14 \\
3 & 110.804773 & -25.807603 & 4.1 & -13.01 & 0.29 & -0.23 & 12 \\
4 & 110.816902 & -25.702447 & 5.6 & -12.69 & 0.88 & -0.04 & 17 \\
5 & 110.703503 & -25.864912 & 6.2 & -13.46 & 0.68 & -0.25 & 8.9 \\
6 & 110.766224 & -25.879853 & 6.9 & -13.27 & 0.03 & -0.31 & 11 \\
7 & 110.615501 & -25.768253 & 6.9 & -13.65 & 0.47 & -0.56 & 19 \\
8 & 110.661126 & -25.674120 & 7.1 & -13.47 & 0.75 & -0.21 & 16 \\
9 & 110.871664 & -25.801926 & 7.2 & -13.36 & 0.62 & -0.24 & 13 \\
10 & 110.814463 & -25.897295 & 8.7 & -13.10 & 0.13 & 0.07 & 12 \\
11 & 110.912240 & -25.742049 & 9.3 & -13.13 & 0.89 & 0.18 & 22 \\
12 & 110.815658 & -25.625267 & 9.4 & -13.67 & 0.15 & -0.19 & 8.7 \\
13 & 110.717331 & -25.591561 & 10.7 & -13.57 & 0.54 & -0.01 & 18 \\
14 & 110.563373 & -25.841295 & 10.7 & -13.27 & 0.53 & 0.13 & 34 \\
15 & 110.768536 & -25.970898 & 12.3 & -12.96 & 0.57 & -0.22 & 18 \\
16 & 110.742678 & -25.973804 & 12.4 & -13.35 & 0.63 & -0.25 & 7.3 
\enddata
\tablenotetext{a}{The offset angle between the X-ray source and the position of VY CMa. }
\tablenotetext{b}{The offset angle between the X-ray source and the nearest 2MASS catalog entry.}
\end{deluxetable*}

\subsection{Constraints on a population of young stars associated with VY CMa} 

We now consider whether the X-ray sources we have detected in the field surrounding VY CMa might represent an associated group of lower-mass (late-type), pre-main sequence stars. 
In Figure~\ref{fighr}, we compare the hardness ratios HR1 (soft \& medium bands) and HR2 (medium \& hard bands) of the 43 strongest X-ray sources to source model predictions. 
We considered an absorbed, solar-abundance, optically-thin thermal plasma model \citep[MEKAL;][]{1985A&AS...62..197M, 1986A&AS...65..511M, 1992Kaastra, 1995ApJ...438L.115L} characterized by an intervening absorbing column $N_H$ and plasma temperature $T_{X}$. 
Our model grid is comprised of 15 logarithmically-spaced $N_H$ values between $10^{20}$ and $3\times10^{23} {\rm ~cm}^{-2}$ and plasma model temperatures between $10^{6}$ to $10^{8}$~K.
This model lacks the sophistication to classify the sources of X-ray emission, but helps establish approximate emission characteristics. 
From Figure~\ref{fighr}, we infer that HR1 is most sensitive to $N_H$, while HR2 is most sensitive to $T_{\rm X}$.

\citet{1978ApJ...219...95L} suggested that VY CMa is associated with the nearby young cluster NGC 2362, based on the photodissociation of molecules in the large molecular cloud complex to the east of VY CMa. 
The eastern half of the XMM observation field of view overlaps with the large molecular cloud complex studied by \citet{1978ApJ...219...95L}, and the cloud appears to have important effects on the detection and characteristics of X-ray sources in our observation. 
First, we note an overall decrease in the X-ray-detected sources towards the cloud (Figure~\ref{fig2}). 
Second, the 16 X-ray sources without IR counterparts (blue squares in Figure~\ref{fig2}) are largely confined to a region where the cloud CO emission is bright \citep[based on comparison with the maps presented in][]{1978ApJ...219...95L} and the extinction is generally higher (Figure~\ref{fig2}). 
Third, the X-ray sources that lack IR counterparts are harder than those with IR counterparts (Figure~\ref{fighr}), suggesting that their soft X-ray emission is heavily attenuated by cloud material. 
Taken together, the evidence therefore indicates many if not most of these X-ray sources without IR counterparts may be young stars that lie embedded in the molecular cloud studied by \citet{1978ApJ...219...95L}, which would place them at the distance of VY CMa.  

The X-ray sources with IR counterparts offer an additional potential constraint on the age of VY CMa. 
Recent parallax distance estimates to NGC 2362 \citep[1.2 kpc;][]{2009MNRAS.400..518M} and VY CMa \citep[$1.2\pm0.1$ kpc;][]{2012ApJ...744...23Z} support the suggestion that VY CMa and the cluster NGC 2362 are associated \citep{1978ApJ...219...95L}. 
NGC 2362 has an age of 5 Myr \citep{2001ApJ...563L..73M}, such that we might expect similarly-aged stars to lie within the field of view of VY CMa, if the hypergiant is associated with the cluster.
To determine if we have detected such a population of young stars, we generated 2MASS color-magnitude and color-color diagrams for the 27 brightest X-ray sources with IR counterparts (Figure~\ref{figcmd}).
In the color-magnitude diagram (upper panel of Figure~\ref{figcmd}) , we have overlaid isochrones of 5 and 10 Myr scaled to the distance of VY CMa (1.3 kpc) and reddened by $A_V\sim1.5 {\rm ~mag}$, which is the typical reddening inferred from extinction maps towards VY CMa \citep[assuming $R_V = 3.1$ and $E_{\rm B-V}$ map from][as shown in Figure~\ref{fig2}]{1998ApJ...500..525S}. 
The placement of the X-ray sources with IR counterparts on these isochrones suggests that a significant fraction of objects that lack measurable proper motions, are brighter than 14 mag in $K_{\rm s}$, and display $J-K_{\rm s}<0.7$ represent a young ($\lesssim 10$ Myr-old) population of late-type stars at the distance of VY CMa. 
The color-color diagram (lower panel of Figure~\ref{figcmd}) further indicates that whereas likely foreground stars (i.e., those with measurable proper motions) lie along the unreddened main-sequence star locus, as expected, most of the 2MASS-detected field sources lie along the locus of reddened main-sequence stars.

The foregoing suggests that at a significant fraction of the $\sim$100 field X-ray sources we have detected in our XMM observations may be representative of a recent epoch of star formation in the vicinity of VY CMa -- possibly the same star formation episode that spawned VY CMa itself. 
A deeper, wider-field X-ray survey of the environment of VY CMa, in combination with deep infrared imaging and spectroscopy, will be required to confirm whether this apparent population of young, X-ray-emitting stars is indeed associated with VY CMa.  
Such a study is warranted, given its potential to place stringent constraints on the age of this enigmatic, mass-losing hypergiant.

\section{Conclusions}

We have obtained XMM-Newton X-ray observations of the red hypergiant VY CMa. We do not detect X-ray emission from the star. 
Using a plasma model with interstellar and circumstellar absorption, we have placed the upper limits on the unabsorbed intrinsic X-ray emission from VY CMa of $F_{\rm X,UL}\approx8\times10^{-14} {\rm ~erg~cm}^{-2} {\rm ~s}^{-1}$ (X-ray flux), $L_{\rm X}<2\times10^{31} {\rm ~erg~s}^{-1}$ (X-ray luminosity), and $L_{\rm X}/L_{\rm bol}< 10^{-8}$ (relative X-ray luminosity).   
These limits allow us to constrain energetic processes that might be responsible for the episodic mass loss observed from VY CMa. 
The limit on (ISM) absorption-corrected $L_{\rm X}/L_{\rm bol}$ ratio of $10^{-9}$ appears to rule out energetic shocks in the wind and ejecta of VY CMa. 
From the upper limit on $L_{\rm X}$, we estimate that the surface-averaged magnetic field strength, $fB$, is less than $10^{-3}$~G. 
Comparison with maser polarization measurements of magnetic field strengths throughout the ejecta of VY CMa, which imply surface magnetic field strengths as large as $\sim200$~G, suggests the filling factor of magnetically active regions on VY CMa ($f$) was $<10^{-5}$ during the epoch of our observation.
However, we stress that our XMM observations represent only a single-epoch snapshot of potential magnetic activity at VY CMa and hence, given the star's record of episodic high mass loss rate events, we may simply be observing the star during a lull between periods of enhanced surface activity.

We have detected more than 100 X-ray emitting sources within the $\sim30^{\prime}$ diameter field of view of the XMM observation. 
The spatial distribution and hardness ratios of these X-ray sources suggest that many are affected by absorption within the molecular cloud associated with VY CMa, thus providing evidence that the sources themselves are at least at the distance of VY CMa.  
We have studied the 43 brightest X-ray sources via their hardness ratios and IR emission, and find that a subset may be late-type pre-MS stars potentially associated with VY CMa.  
The near-IR magnitudes and colors of roughly half of the 18 X-ray sources with IR counterparts are consistent with these stars being a young ($\lesssim$10 Myr-old) population that is similar in age and distance to the nearby cluster NGC 2362.  
Further study of this apparent group of young stars that is potentially associated with VY CMa may help constrain the age of the hypergiant itself.

\acknowledgments

{\it Facilities:} \facility{XMM (EPIC)}.

We acknowledge support from NASA Award Number, NNX13AB55G for our XMM observations of VY CMa.
This research has made use of the NASA/ IPAC Infrared Science Archive, which is operated by the Jet Propulsion Laboratory, California Institute of Technology, under contract with the National Aeronautics and Space Administration.
This publication makes use of data products from the Two Micron All Sky Survey, which is a joint project of the University of Massachusetts and the Infrared Processing and Analysis Center/California Institute of Technology, funded by the National Aeronautics and Space Administration and the National Science Foundation.

\end{document}